\documentclass[aps,prb,amsmath,twocolumn,superscriptaddress,showpacs,floatfix]{revtex4-1}
\usepackage{graphicx}
\usepackage{xcolor}
\usepackage[version=4]{mhchem}
\usepackage[colorlinks=true,urlcolor=blue,anchorcolor=blue,linkcolor=blue,citecolor=blue,breaklinks=true]{hyperref}
\usepackage[normalem]{ulem}
\usepackage{physics}
\newcommand{\RN}[1]{%
  \textup{\uppercase\expandafter{\romannumeral#1}}%
}

\newcommand{\etal}{{\it et al.}}

\definecolor{applegreen}{rgb}{0.55, 0.71, 0.0}

\begin{document}

\title{Effects of spin orbit coupling in superconducting proximity devices -- application to \ce{CoSi$_2$}/\ce{TiSi$_2$} heterostructures}

\author{Vivek Mishra}
\affiliation{Kavli Institute for Theoretical Sciences, University of Chinese Academy of Sciences, Beijing 100190, China}

\author{Yu Li}
\affiliation{Kavli Institute for Theoretical Sciences, University of Chinese Academy of Sciences, Beijing 100190, China}

\author{Fu-Chun Zhang}
\email{fuchun@ucas.ac.cn}
\affiliation{Kavli Institute for Theoretical Sciences, University of Chinese Academy of Sciences, Beijing 100190, China}
\affiliation{CAS Center for Excellence in Topological Quantum Computation, University of Chinese Academy of Sciences, Beijing 100190, China}

\author{Stefan Kirchner}
\email{stefan.kirchner@correlated-matter.com}
\affiliation{Zhejiang Institute of Modern Physics \& Department of Physics, Zhejiang University, Hangzhou 310027, China}
\affiliation{Zhejiang Province Key Laboratory of Quantum Technology and Device, Zhejiang University, Hangzhou 310027, China}

\date{\today}

\begin{abstract}
Motivated by the recent findings of unconventional superconductivity in \ce{CoSi2}/\ce{TiSi2} heterostructures, we study the effect of interface induced Rashba spin orbit coupling on the conductance of a three terminal ``T" shape superconducting device. We calculate the differential conductance for this device within the quasi-classical formalism that includes the mixing of triplet-singlet pairing due to the Rashba spin orbit coupling. We discuss our result in the light of the conductance spectra reported by Chiu \etal for \ce{CoSi2}/\ce{TiSi2} heterostructures. 
\end{abstract}
%\pacs{}
\maketitle

\textit{Introduction--}
The search for platforms that can host Majorana zero modes (MZMs) has been one of the major topics driving current condensed matter research as  MZMs, being localized quasiparticles that obey non-Abelian braiding statistics,  are the essential ingredient for  topological quantum computing\cite{ReadGreen,Kitaev2001,Ivanov2001,Kitaev2003,Wilczek2009,DasSarmaRMP}.  
Early proposals for creating MZMs involve spin triplet superconductivity while almost all known superconductors belong to the spin singlet class with a few possible exceptions such as \ce{UPt3}. 
As a result,  a variety of ingenious heterostructures of superconducting nano-wires have been proposed and observed to generate the required $p\mbox{-}$wave pairing component, taking advantage of broken  time-reversal and inversion symmetry\cite{PW1,PW2,Mourik1003,Nadj-Perge602}. 
MZM has also been proposed inside a vortex of topological superconductor \cite{VolovikBook}; the experimental observation of MZM in some of the iron-based superconductors is along this line\cite{HongDingWang333,DLFeng2018}.  Fu and Kane \cite{FuKane2008} proposed that proximity of s-wave superconductor on surface of 3D topological insulator may serve for the same purpose to generate MZMs, due to the spin-momentum locking.  Their proposal has been confirmed experimentally\cite{JinFeng2016}. 

\begin{figure}
\includegraphics[width=.88\linewidth]{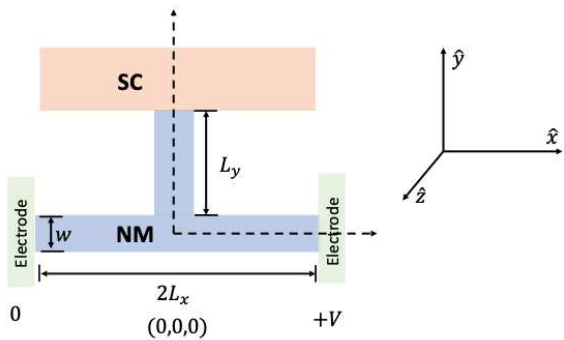}
\caption{ Schematic illustration of a T-shaped junction. The three terminal proximity device consists of a diffusive normal metal (NM) part attached to a superconductor (SC). The blue area indicates the diffusive normal metal (NM) part of the device, while the orange area  shows the SC part  (color online).}
\label{Fig:Tshape}
\end{figure}

To distinguish between singlet and triplet superconductors, in addition to nuclear magnetic resonance\cite{NMR_RMP} and $\mu$-spin rotation probe, a  T-shaped proximity structure junction was proposed to  probe into the presence of triplet superconductivity  \cite{Asano2007}. The proposed device consists of two normal metal wires combined to form the  letter `T'. This three-terminal device is connected to a superconductor at the free end of the leg, see Fig.\ \ref{Fig:Tshape}. As shown in Ref.\ \cite{Asano2007},
a chiral $p\mbox{-}$wave or an ordinary $p\mbox{-}$wave state give a zero bias conductance peak (ZBCP)  in response to a bias voltage between the open ends of the bar of the `T'. 

Recent experimental results reported by Chiu \textit{et al.} have been argued to be consistent with the occurrence of  chiral $p\mbox{-}$wave pairing in \ce{CoSi$_2$}/\ce{TiSi$_2$} heterostructures \cite{Chiu2020}. Chiu \textit{et al.} support their claim with conductance spectroscopy data of   \ce{CoSi$_2$}/\ce{TiSi$_2$} superconductor-normal metal (SN) tunnel junctions. In these heterostructures, \ce{CoSi$_2$} is the superconducting component which becomes superconducting below $1.5~$K. The conductivity of the SN tunnel junctions agrees with the theoretical calculations based on the Blonder-Tinkham-Klapwijk (BTK)  model for a chiral $p\mbox{-}$wave superconductor\cite{BTK}. However, there is a sharp zero bias peak in the conductance spectra of the SN junction, which cannot be described within the BTK theory.

Chiu \textit{et al.} further substantiate  their interpretation with conductance spectra based on three terminal T-shaped proximity devices  similar to the one sketched in  Fig. \ref{Fig:Tshape},
which again show ZBCPs. As noted in Ref.\ \cite{Chiu2020}, a distinction between ordinary and chiral $p\mbox{-}$wave superconductor  solely based on experimental conductance spectra is hardly feasible.
The observation of hysteresis behavior in the magnetoresistance below  superconducting transition temperature $T_c$ of the \ce{CoSi$_2$}/\ce{TiSi$_2$} junctions, however, further vindicates their claim of chiral $p\mbox{-}$wave in the \ce{CoSi$_2$}/\ce{TiSi$_2$} heterostructures. 
The findings of Ref. \onlinecite{Chiu2020} are  intriguing for a few reasons. The superconductivity in \ce{CoSi$_2$} was discovered in 1952\cite{CoSi2SC}, the theoretical estimate of $T_c$ based on phonon mediated pairing appears to agree well with the experimental $T_c$\cite{CoSi2ES}. The specific heat data below superconducting state suggests conventional $s\mbox{-}$wave pairing\cite{CoSi2SH}. This material does not appear to be located in the vicinity of magnetism, therefore there is no reason to expect it to be a chiral $p\mbox{-}$wave superconductor, at least in the bulk limit. 

It is worth noting that a strong spin-orbit coupling (SOC), exceeding the superconducting gap of \ce{CoSi$_2$} by more than a factor 30,  has been reported in \ce{CoSi$_2$} by the same group\cite{Chiu2020}. Having in mind this strong SOC, we propose the substrate induced Rashba SOC as a source of $p\mbox{-}$wave pairing in this system. It is known that SOC induced pairing does not break the time reversal symmetry\cite{GorkovRashba}, and the presence of the SOC also leads to mixing of the triplet and the singlet components. 
In the context of noncentrosymmetric superconductors, the tunneling conductance for SN junctions has been studied in systems with SOC\cite{Iniotakis2007,VorontsovABS,Tanaka2010,EschrigABSNCS,TamuraProximityRashba}. However, the effect of the SOC and the conductance for a mixed parity superconductor in a T-junction device is not known. 
The original T-junction study did not include the SOC, and the ZBCP for a chiral $p\mbox{-}$wave superconductor is expected to be weak\cite{Asano2007}.

In this paper, we investigate the effect of the singlet-triplet mixing on the conductance spectra of  the  T-shaped junctions.  In the context of \ce{CoSi$_2$}/\ce{TiSi$_2$} heterostructures, we focus on substrate induced SOC in a superconductor, which results in the ``$sp$'' pairing state, where the singlet component has  $s\mbox{-}$wave symmetry and the triplet component has  $p\mbox{-}$wave symmetry. 
In principle,  triplet and  singlet components can have anisotropic structures due to the orbital form factors and because the bands crossing the Fermi energy derive from the 3d orbitals of \ce{Co} \cite{CoSi2ES}. Thus,  we also consider the ``$df$'' and ``$dp$'' pairing states, which have additional orbital form factors compatible with $d_{x^2-y^2}$ and $d_{xy}$ functions, which lead to $d_{x^2-y^2}$ and $d_{xy}$ structures for the singlet components and effectively $f$-wave and $p$-wave like structures for the triplet components, respectively.
The $df$ state has been proposed for a few heavy electron noncentrosymmetric systems\cite{Tada2008,Yanase2008} and the $dp$ state  has been suggested for  \ce{LaAlO$_3$}/\ce{SrTiO$_3$} heterointerfaces\cite{Tanaka2010}. 

%%%%%%%%%%%%%%%
%%%%%%%%%%%%%%%

\textit{Model} \& \textit{Formalism--} We model the  \ce{CoSi$_2$}/\ce{TiSi$_2$} T-shaped junction of Ref.\cite{Chiu2020} in terms of  the two-dimensional proximity devices depicted schematically in Fig. \ref{Fig:Tshape}. The transport in the normal metal (NM) part is assumed to be  diffusive  which is the experimentally relevant regime.
The height ($d$) and the width ($w$) are very small compared to its length ($L_{x/y}$) in either direction and its dimensions are assumed to be  very small compared to the coherence length $\xi_0\equiv \hbar v_F/\pi \Delta$ (i.e., $w,d \ll \xi_0$), where $v_F$ and $\Delta$ are the Fermi velocity and the superconductivity gap, respectively. Within the ambit of these assumptions,  this structure can be thought of a set of  two one dimensional wires joined to form the shape of  the letter `\textbf{T}'. The leg of this T-shaped junction is attached to a clean superconductor. The ends of the horizontal section of this junction are subjected to a bias voltage ($eV$). We consider the case where the SOC exists in the superconducting component of this structure due to its  broken inversion symmetry.\\ 
The kinetic part of the Hamiltonian reads,
\begin{eqnarray}
H_{\bf{k}} &=& \xi_{\bf{k}} + H_{SOC,\bf{k}}.
\label{eq:Hm}
\end{eqnarray}
Here $\xi_{\bf{k}}$ is the electronic dispersion relation for the fermions and  the SOC term is,
\begin{eqnarray}
H_{SOC,\bf{k}} &=& \alpha \left( \boldsymbol{\sigma} \times \bf{k}\right)\cdot \hat{z} =\alpha \boldsymbol{\mathcal{A}}_{\bf{k}}\cdot \boldsymbol{\sigma}, 
\label{eq:HMSOC}
\end{eqnarray} 
where $\alpha$ is the Rashba SOC coupling constant, $m$ is the effective mass and $\boldsymbol{\sigma}$ is ($\sigma_x,\sigma_y,\sigma_z$), where $\sigma_{x/y/z}$ are the Pauli matrices in spin space. 
We consider the Rashba SOC, that is induced along the growth direction, which is chosen to be the $\hat{z}$ direction in the T-shaped junction.  In this case, the normal to the relevant interface is along the $\hat{z}$ axis and the SOC vector is,
\begin{equation}
\boldsymbol{\mathcal{A}}_{\bf{k}} = \left( k_y,-k_x ,0\right) = \abs{k}(\sin \phi_k , -\cos \phi_k ,0).
\label{eq:ASOC}
\end{equation}
where $\phi_k$ is the angle in the two dimensional momentum space.

Diagonalizing the Hamiltonian results in a splitting of the original band into two  helical bands with different spin structures. 
The energies of these two bands are $\xi_{\bf{k}}\pm \alpha \abs{\bf{k}}$. The difference in the density of states and the Fermi velocities are of the order of $\alpha p_F/E_F$, where $p_F$ and $E_F$ are the Fermi momentum and the Fermi energy of the original band. For realistic systems, the SOC energy is generally very small compared to the Fermi energy. Therefore, we ignore this difference in the density of states and the Fermi velocities between the helical bands and we take these parameters to be the same as the original band for our subsequent  calculations. 

We assume that the superconducting component is confined in the two-dimensional plane and that it has dimensions that are very large compared to the coherence length, hence we treat it like a homogeneous system, and ignore any kind of inverse proximity effect due to the junction formation. We adopt the quasi-classical Keldysh formalism to carry out the conductance calculations\cite{Usadel1970}, where the quasi-classical Green's function consists of retarded, advanced and Keldysh components. Each of these components is a $4\times4$ matrix in the Nambu-spin space. We denote the $4\times4$ Green's function in this space with $\check{...}$ and $\hat{...}$ denotes the $2\times2$ Green's functions in the spin basis. The advanced and Keldysh components can be obtained from the retarded component, on which we focus 
in the following. Following Ref. \onlinecite{Hayashi2006}, the quasi-classical retarded Green's function in a superconductor without inversion symmetry can be expressed as,
\begin{eqnarray}
\check{g} &=& \begin{pmatrix}
g_{\RN{1}} \sigma_{\RN{1}} +g_{\RN{2}} \sigma_{\RN{2}}  & -\left( f_{\RN{1}} \sigma_{\RN{1}} +f_{\RN{2}} \sigma_{\RN{2}}\right)i\sigma_y \\
-i\sigma_y \left( \bar{f}_{\RN{1}} \sigma_{\RN{1}} +\bar{f}_{\RN{2}} \sigma_{\RN{2}}\right)&  \sigma_y \left( \bar{g}_{\RN{1}} \sigma_{\RN{1}} +\bar{g}_{\RN{2}} \sigma_{\RN{2}} \right) \sigma_y
\end{pmatrix},
\label{Eq:QGSC}
\end{eqnarray}
where $\sigma_{\RN{1}/\RN{2}}=( \sigma_0 \pm  \boldsymbol{\mathcal{A}}_{\bf{k}}\cdot \boldsymbol{\sigma})/2$, $g_{\RN{1}/\RN{2}}(\varepsilon) = {\varepsilon}/{\sqrt{\varepsilon^2-\Delta_{\RN{1}/\RN{2}}^2}}$,  $f_{\RN{1}/\RN{2}}(\varepsilon) = {\Delta_{\RN{1}/\RN{2}}}/{\sqrt{\varepsilon^2  -\Delta_{\RN{1}/\RN{2}}^2}}$, $\bar{g}_{\RN{1}/\RN{2}}=-{g}_{\RN{1}/\RN{2}}$ and $\bar{f}_{\RN{1}/\RN{2}}={f}_{\RN{1}/\RN{2}}$.  The general gap structure for a system with the SOC is\cite{Frigeri2004,*FrigeriArxiv},
\begin{eqnarray}
\hat{\Delta} = \left( \Delta_{s} \Phi_s ( \phi_k) + \Delta_{t} \Phi_t (\phi_k) \boldsymbol{\mathcal{A}}_{\bf{k}}\cdot \boldsymbol{\sigma} \right) i\sigma_y.
\label{eq:GS1}
\end{eqnarray}
 Here the SOC vector $\boldsymbol{\mathcal{A}}_{\bf{k}}$ acts like the $\mathbf{d}\mbox{-}$vector, and $\Delta_{s}$ ($\Delta_{t}$) is the gap magnitude of the singlet (triplet) component. The gaps on two helical bands are $\Delta_{\RN{1}/\RN{2}}=\Delta_s \Phi_s \pm \Delta_t \Phi_t$.  The angular anisotropy of the gaps are embedded in $\Phi_s$ and $\Phi_t$. The simplest case is $\Phi_s =\Phi_t = 1$, which is referred as $sp\mbox{-}$state, where the singlet component is an isotropic $s\mbox{-}$wave state, and the triplet component has $p\mbox{-}$wave structure. Such states have been proposed for various non-centrosymmetric superconductors. Apart from this, the other possibilities are the $df\mbox{-}$state, where $\Phi_s = \Phi_t = \cos 2\phi_k $, and the $dp\mbox{-}$state with $\Phi_s=\Phi_t=\sin 2\phi_k$.  We focus on $sp\mbox{-}$state, which is more relevant in the context of \ce{CoSi$_2$}/\ce{TiSi$_2$} heterostructures.
The gap function is parameterized as
\begin{eqnarray}
\hat{\Delta} = \Delta_0 \left( \frac{ 1}{\sqrt{1+r^2}} + \frac{r }{\sqrt{1+r^2}} \boldsymbol{\mathcal{A}}_{\bf{k}}\cdot \boldsymbol{\sigma}\right)i\sigma_y.
\end{eqnarray}
where $\Delta_0$ is $\sqrt{\Delta_s^2+\Delta_t^2}$\cite{Annunziata2012}. Here the parameter $r\in [0,\infty]$ is the ratio of triplet to singlet component.

%%%%%%%%%%%%%%%
\begin{figure*}
\includegraphics[width=.325\linewidth]{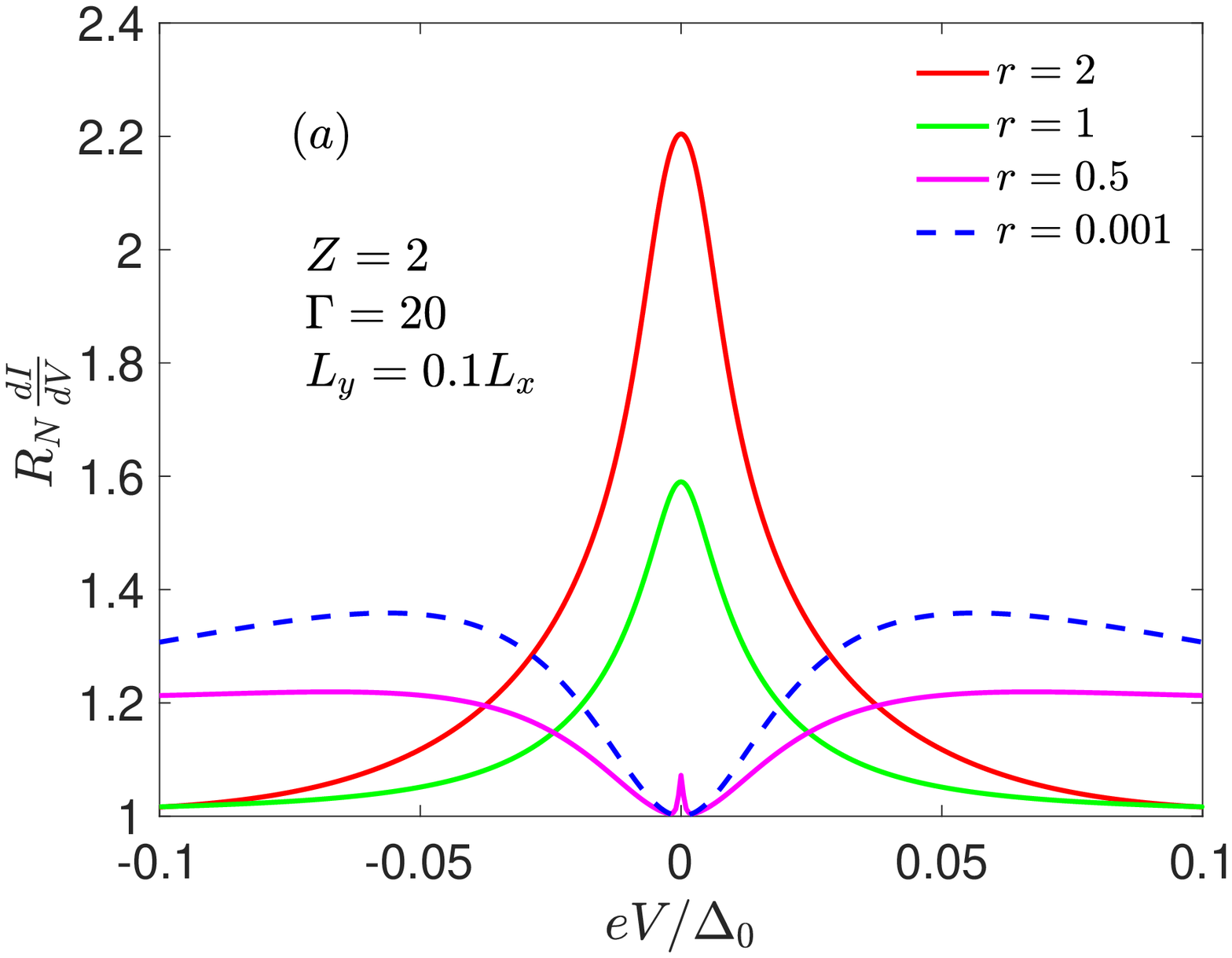}
\includegraphics[width=.325\linewidth]{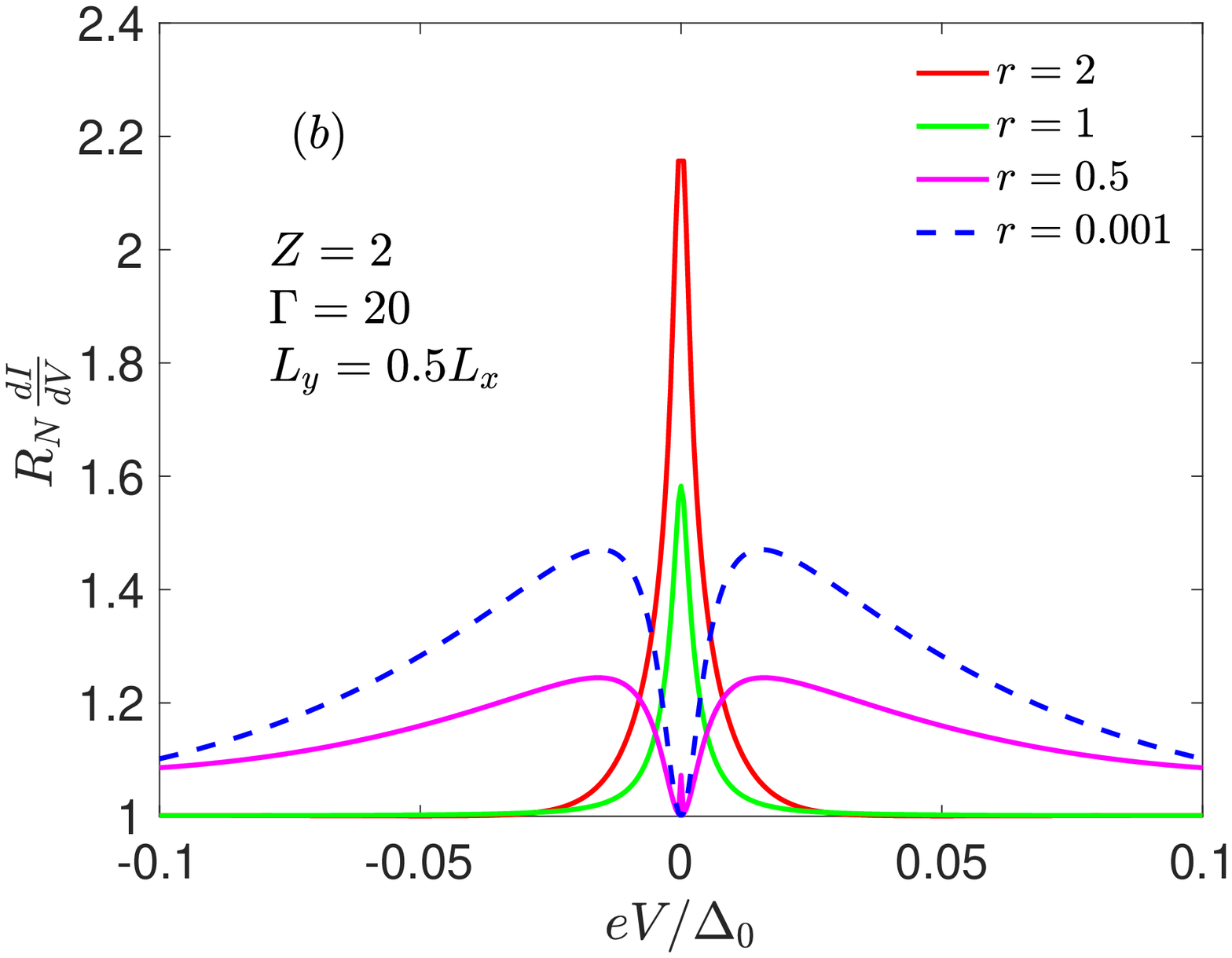}
\includegraphics[width=.325\linewidth]{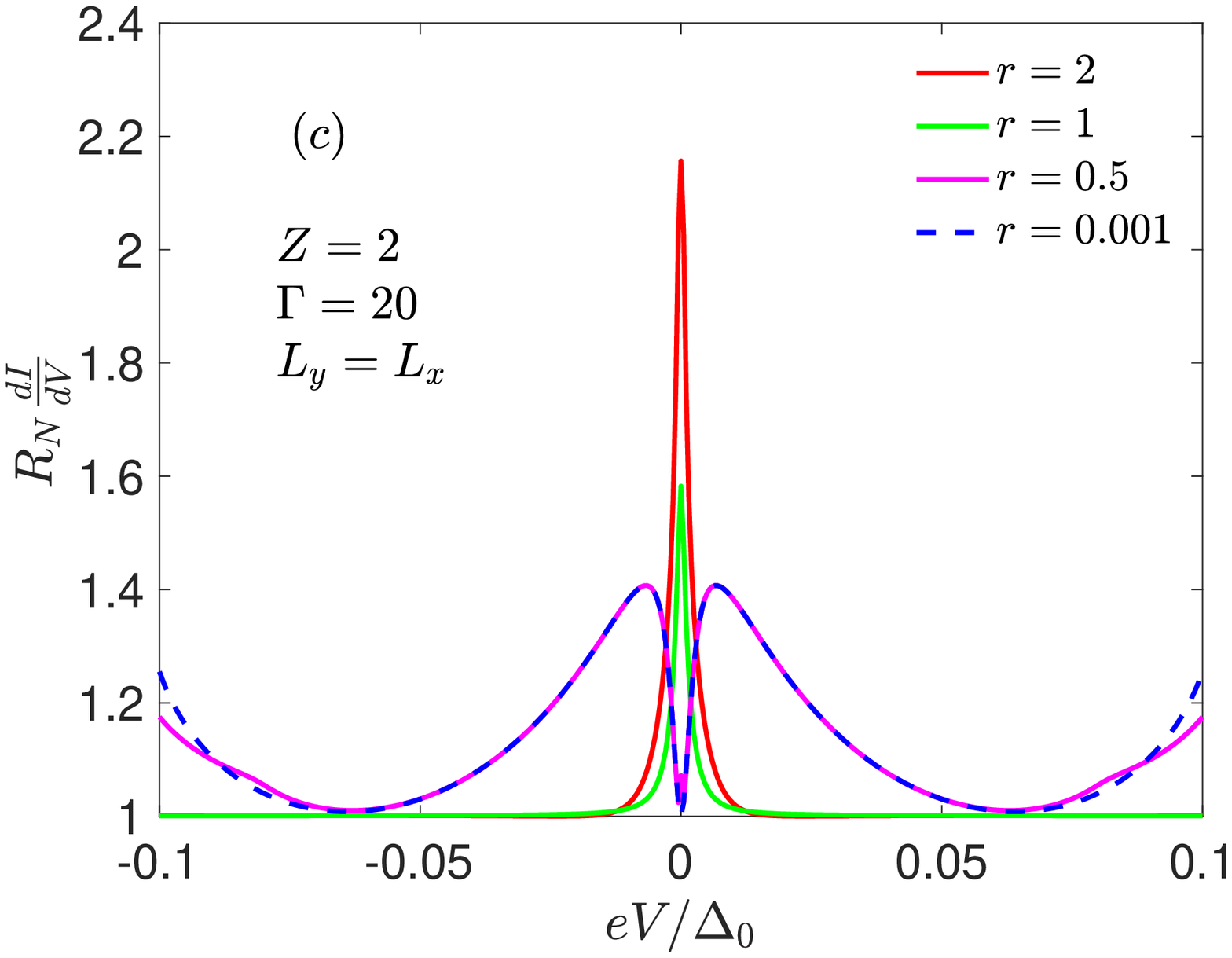}
\caption{Differential conductance for a T-shaped junction attached to a  superconductor under the influence of Rashba SOC. The parameter $r$ is the ratio of triplet to  singlet component of the order parameter. The interface quality parameter $\Gamma$ is set to $20$. The ratio of lengths along the two spatial direction $L_y/L_x$ is $0.1$, $0.5$ and $1$ in the panels $(a)$, $(b)$ and $(c)$, respectively. The differential conductance is calculated in the zero temperature limit.}
\label{Fig:didv_rv}
\end{figure*}
%%%%%%%%%%%%%

The Cooper pairs from the superconducting side can  tunnel into the diffusive normal metal (NM), and this effect is included through the  boundary conditions, which are used to solve the Usadel equations on the NM  side. We treat the barrier between the NM  and the superconductor as a spin-independent barrier. This assumption is justified because the SOC is very small compared to the Fermi energy\cite{EschrigABSNCS}.  We first calculate the retarded component of the quasi-classical Green's function $\check{g}^R_n$, and then construct the advanced and the Keldysh components using $\check{g}^R_n$. The subscript $n$ denotes the normal metal. The Usadel equations for $\check{g}^R_n$ are
 \begin{eqnarray}
 D \partial_\ell (\check{g}^R_n \partial_\ell \check{g}^R_n) +i \left[ \varepsilon\check{\tau_3},\check{g}^R_n\right] &=&0.
\end{eqnarray} 
 where $D$ is the diffusion constant of the normal metal, $\ell$ denotes the spatial directions $x/y$,  and  $\check{\tau}_3 $ is $\mathrm{diag}(1,1,-1,-1)$. The normalization condition for the quasi-classical Green's function is,
 \begin{eqnarray}
 \check{g}^R_n\check{g}^R_n = \check{\mathbf{1}}.
 \label{eq:norm}
 \end{eqnarray}
These equations are supplemented by the boundary conditions,
\begin{eqnarray}
\check{g}^{R}_n &=& \check{\tau}_3, \\
 \check{g}^R_n\nabla\partial_y\check{g}^R_n &=& \check{0},
\end{eqnarray}
where the last condition reflects  current conservation\cite{Zaitsev1994}. The boundary condition at $(0,L_y)$ depends on the nature of the gap in the superconductor.  We use the boundary condition derived by Nazarov for interfaces with arbitrary transparency\cite{Nazarov1999}, which was generalized for unconventional superconductors by Tanaka \textit{et al}\cite{Tanaka2003bc,Tanaka2004,*Tanaka2004_ERR,Tanaka2005,*Tanaka2005_ERR}. In this approach, the interface is modeled as a $\delta$ function potential barrier $H\delta(y-L_y)$, which has the transmission probability,
\begin{eqnarray}
T(\phi) = \frac{4 \cos^2 \phi}{4\cos^2 \phi +Z^2},
\label{eq:Tphi}
\end{eqnarray}
where $\phi$ is the angle measured with respect to the normal to the interface, which is $y$ axis in the geometry we consider, and $Z$ is a dimensionless parameter given by $Z=2mH/k_F^2$. Here $m$ is the effective mass and $k_F$ is the Fermi momentum. A large value of $Z$ gives an interface with poor transparency, whereas $Z=0$ characterizes a transparent interface. The boundary condition at the SN interface can be expressed as,
\begin{eqnarray}
 L_y \check{g}_n \frac{\partial \check{g}_n}{\partial y}\Big|_{y=L_y} = 2 \Gamma \left\langle [\check{g}_n, \check{B}(\phi)] \right\rangle_\phi,
\label{Eq:BC1}
\end{eqnarray}
where $\check{g}_n$ at the right hand side of Eq. \eqref{Eq:BC1} is the quasi-classical Green's function in the NM region evaluated at $(0,L_y)$, and $\Gamma$ is the ratio of  the  normal metal  resistance $R_N$, and the interface resistance $R_B$.
The angular average at the right hand side of Eq.\eqref{Eq:BC1} is defined as,
\begin{eqnarray}
\left\langle ... \right\rangle_{\phi} = \frac{\int_{-\pi/2}^{\pi/2} d\phi (...) \cos \phi}{\int_{-\pi/2}^{\pi/2} d\phi T(\phi) \cos \phi}
\end{eqnarray}
Note, the angle $\phi$ in the boundary condition is measured with respect to the interface normal. The matrix function $\check{B}$ in Eq. \eqref{Eq:BC1} is,
\begin{eqnarray}
\check{B}(\phi) &=& \left[ -T^{\prime} [\check{g}_n, \check{H}^{-1}_{-}]+ \check{H}^{-1}_{-} \check{H}_+ - T^{\prime 2} \check{g}_n \check{H}^{-1}_{-} \check{H}_+\right]^{-1} \nonumber \\
&\times& \left[ -T^{\prime} (\check{\mathbf{1}}+\check{H}^{-1}_{-})+ T^{\prime 2} \check{g}_n \check{H}^{-1}_{-} \check{H}_+  \right], \\
T^{\prime}(\phi) &=& \frac{T(\phi)}{2-T(\phi) + 2\sqrt{1-T(\phi)}},\\
\check{H}_{\pm} &=& \frac{1}{2} \left[ \check{g}(\phi)\pm\check{g}(\pi-\phi) \right].
\end{eqnarray}
Note, the boundary condition itself depends on the solution at the boundary. 
To calculate the differential conductance, we first calculate the current. For the current calculation, we need the Keldysh component  of the quasi-classical Green's function,
\begin{equation}
\check{g}^K = \check{g}^R \check{h} - \check{h} \check{g}^A,
\end{equation}
where the advanced component is,
\begin{eqnarray}
\check{g}^A = - \check{\tau}_3 (\check{g}^R)^\dagger \check{\tau}_3,
\end{eqnarray}
and the spin resolved distribution function $\check{h}$ is a diagonal matrix $\mathrm{diag}(f_{L\uparrow}+f_{T\uparrow},f_{L\downarrow}+f_{T\downarrow},f_{L\uparrow}-f_{T\uparrow},f_{L\downarrow}-f_{T\downarrow})$\cite{Morten2005}, where $f_{T\nu}$ and $f_{L\nu}$ are the transverse and the longitudinal distribution functions and $\nu$ is the spin index.
In  the T-shaped junction, a bias voltage ($V$) is applied at $x=+ L_x$ and at the other end the voltage is kept at zero. Therefore, the equilibrium spin-resolved distribution functions at these two ends are,
\begin{eqnarray}
f_{T_{\uparrow/\downarrow}}\Big|_{x=L_{x},y=0} &=&\frac{1}{2}\left[ n_f(\frac{\varepsilon_-}{2T})- n_f(\frac{\varepsilon_+}{2T}) \right],
 \label{eq:ftc1} \\
f_{T_{\uparrow/\downarrow}}\Big|_{x=-L_{x},y=0} &=& 0.
\label{eq:ftc2}
\end{eqnarray}
Here $n_f$ is the Fermi-Dirac distribution function and $\varepsilon_\pm = \varepsilon\pm eV$. The transverse component of the distribution function will be the same for both spin components at the normal electrode.  The charge current density  is,
\begin{equation}
J_E (x,T) = \frac{e N_0 D}{8} \int_{-\infty}^{\infty} d\varepsilon \mathrm{Tr} \left[ \hat{\tau}_3 ( \check{g}^R \partial_x \check{g}^K +\check{g}^K \partial_x \check{g}^A  ) \right].
\end{equation}
 Here $N_0$ is the total density of states at the Fermi level. The differential conductance can be obtained by evaluating the derivative of the charge current density {\itshape w.r.t.} the bias voltage -- we numerically solve the Usadel equations in the normal metal and with   the aforementioned boundary conditions. Since the boundary condition at the SN interface involves the solutions at the interface, so we start with a guess solution and obtain the final solution self-consistently.   
%%%%%

\textit{Results} \& \textit{Discussion--} We consider a good interface between NM and SC and fix $\Gamma$  at a value of $20$. The interface barrier parameter $Z$ is set to $2$.  A larger value of $\Gamma$ represents a good quality surface, which is essential for the formation of a sizable proximity effect.  Figure \ref{Fig:didv_rv} shows the differential conductance for the T-shaped device, where the superconducting portion is under the influence of the substrate induced Rashba SOC for several values of the parameter $r$ indicating the relative strength of the triplet component.  The magnitude of the gap is $0.05E_{th}$, where $E_{th}$ is the Thouless energy for the half wire along the $\hat{x}$ direction {\itshape i.e.} $E_{th} \equiv \hbar D /L_x^2$. For the proximity problem, the characteristic energy scale in the NM  is the Thouless energy,  which is inversely proportional to the square of the device length. A smaller devices is usually better for observing proximity effect related physics.  

\begin{figure}
\includegraphics[width=.97\linewidth]{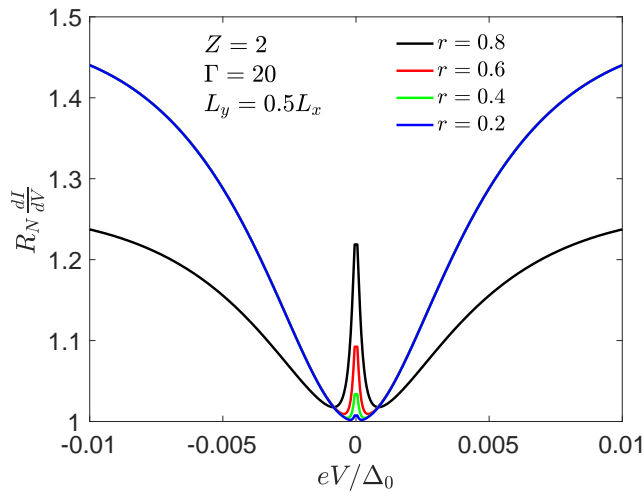}
\caption{The differential conductance for a T-shaped junction of a $dp\mbox{-}$superconductor for several values of $r$ with larger singlet component. The length along the $\hat{y}$ direction is $0.5L_y$. }
\label{Fig:rint}
\end{figure}

For large values of $r$, the triplet component dominates. In this regime  we find that the differential conductance is similar to the $p$-wave case\cite{Asano2007}. In general, for a three dimensional system, a $\hat{z}$ Rashba SOC gives $\Delta_s \pm \Delta_t \sin \theta$, where $\theta$ is the polar angle. Therefore, a triplet dominated system will have horizontal line nodes. However, in our study we consider a two-dimensional system, where the gaps on the two helical bands are $\Delta_s \pm \Delta_t$. Thus,  in the triplet dominant limit ($r>1$), we have isotropic unequal gaps on two bands with opposite chiralities. A zero bias conductance peak (ZBCP) is expected for a chiral $p$-wave superconductor\cite{Asano2007}, however it is expected to be weaker than a $p\mbox{-}$wave system. We find that the height of the peak is comparable to that of a $p\mbox{-}$wave system. Unlike in a chiral $p\mbox{-}$wave superconductors,  the time-reversal symmetry is  not broken in the case considered here.  The origin of the peak is the symmetry of the induced pairing in NM. In the diffusive metal,  the isotropic $s\mbox{-}$wave state can survive due to impurity scattering, which kills any other kind of superconducting state. In the superconducting side of the junction, both triplet and singlet components are even functions of frequency. Therefore, the triplet component leaks odd frequency, even parity and spin triplet pairs, and the odd frequency nature of these induced pair gives rise to a ZBCP\cite{TanakaTriplet,*TanakaTriplet_ERR,TanakaGolubov2007}. In the case of two helical bands with opposite chirality triplet state, the spectral weight of the ZBCP is larger than that expected for a chiral superconductor. The ZBCP becomes sharper as the length of the leg ($L_y$) attached to the superconductor increases, therefore a T-shaped junction with a shorter leg provides a better chance of ZBCP detection. We find that a  ZBCP forms as long as the triplet component is stronger.  For the special case of $r=1$, when triplet and singlet components are equal, we still find a ZBCP in the differential conductance albeit with a reduced height and width. Since one of the bands has a zero gap in this limit, the height of the peak decreases -- the origin of the reduced width in the ZBCP for smaller $r$ is the presence of even frequency, spin singlet and even parity pairs, which comes from the singlet component in the mixed parity superconducting state. Such pairs reduce the density of states at the Fermi level, which reduces the conductivity. However, induced pairs also increase conductivity in the diffusive metal, this increase comes through Maki-Thompson like process\cite{VolkovTakayanagi,*VolkovTakayanagi2}. These two counter effects cancel at  zero energy. The finite energy maximum  in the conductivity near the Thouless energy scale arises due to different decay patterns of these two effects. The negative contribution from the loss of density of states decays exponentially, while the Maki-Thompson like contribution decays non-exponentially over the energy scale of $E_{th}$. These two opposite contributions result in a dip in the limit of pure singlet superconductor ($r\ll 1$) in the T-shaped junction. In the singlet dominated regime ($0<r<1$), we find both a dip from the singlet component and a weak ZBCP from the triplet component. Figure \ref{Fig:rint} shows evolution of the conductance peak  to a dip in the strong singlet limit. The width and height of the ZBCP decreases rapidly with diminishing triplet component.

%%%%%%%%%%%%%%%
%%%%%%%%%%%%%%%
\begin{figure}
\includegraphics[width=.97\linewidth]{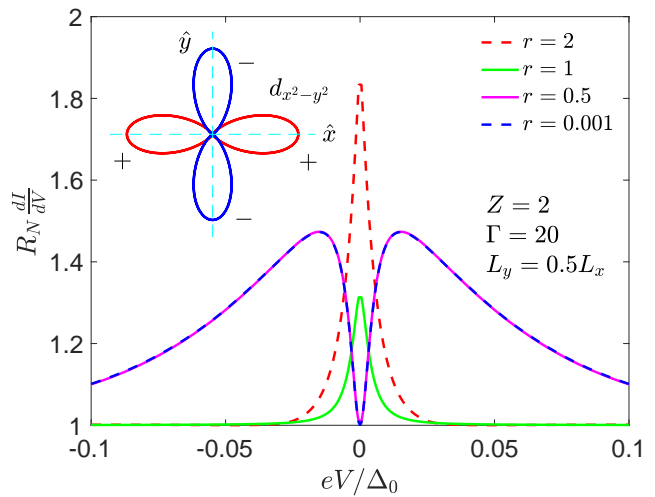}
\caption{The differential conductance for a T-shaped junction of a $df\mbox{-}$superconductor for several values of $r$. The length along the $\hat{y}$ direction is $0.5L_y$. The orientation of $d_{x^2-y^2}$ orbital form factor is shown in the main figure. The leg of the T-junction is taken along the $\hat{y}$ direction, and the voltage is applied along the $\hat{x}$ direction.}
\label{Fig:c_df}
\end{figure}
%%%%%%%%%%%%%

Next, we consider $df$ and $dp$ states  which possess anisotropic orbital components. For the  $df\mbox{-}$state, $\Phi_s$ and $\Phi_t$ are modeled by $\cos 2\phi_k$. For this gap function, there are two line nodes at an angle $\pm \pi/4$ {\itshape w.r.t.} the interface normal $\hat{y}$ axis. Since we have already shown that a T-shaped junction with shorter leg length is better for observing the ZBCP,  we fix the value of $L_y$ at $0.5L_x$ and consider a good quality interface with $\Gamma=20$ and $Z=2$ for our differential conductance calculations with anisotropic form factors.  
Fig. \ref{Fig:c_df} shows the differential conductance for a T-shaped junction attached to $df\mbox{-}$symmetry superconductor. We find qualitatively similar behavior for a $df\mbox{-}$superconductor to that of a $sp\mbox{-}$superconductor which was discussed above. This qualitative similarity between $sp$ and $df$ superconductors can be understood by examining the phase shift in the  gap functions of incoming and outgoing quasiparticle trajectories at the interface. For the $d_{x^2-y^2}$ orbital function, the incoming  $\Delta(\phi)$ and the outgoing $\Delta(\pi-\phi)$ are qualitatively the  same as $sp\mbox{-}$ superconductor.  The nodal line of the $d_{x^2-y^2}$ form factor  is at an angle of $\pm \pi/4$, so there is no additional sign change due to this anisotropic factor, and the triplet component is effectively the same as for the $sp\mbox{-}$ superconductor.  However, the $d_{x^2-y^2}$ form factor reduces the height of the ZBCP in the triplet dominated regime ($r\geq 1$), and in the strong singlet regime $r<1$, the tiny peak that we find for the $sp$ superconductors is smeared and the lineshape is similar to a $s\mbox{-}$wave superconductor. 
\begin{figure}
\includegraphics[width=.97\linewidth]{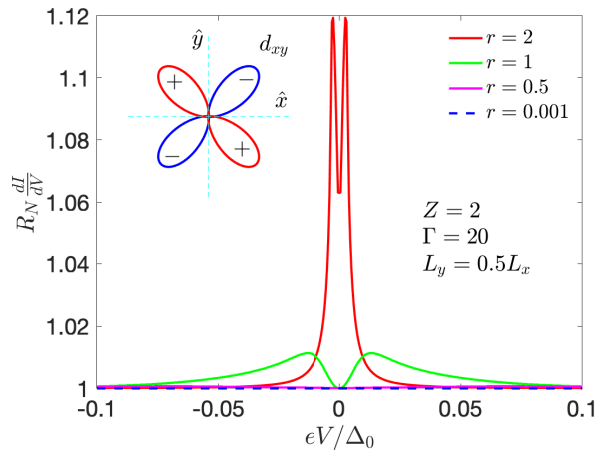}
\caption{The differential conductance for a T-shaped junction of a $dp\mbox{-}$superconductor for several values of $r$. The length along the $\hat{y}$ direction is $0.5L_y$. The orientation of $d_{xy}$ orbital form factor is shown in the main figure. The leg of the T-junction is taken along the $\hat{y}$ direction, and the voltage is applied along the $\hat{x}$ direction.}
\label{Fig:c_dp}
\end{figure}

In contrast, we find a qualitatively different behavior for the $dp$ state, as shown in the Fig. \ref{Fig:c_dp}, there is a splitting of the ZBCP with reduced heights. For a $d_{xy}$ orbital factor, the nodal line is along the interface normal, therefore for all the incoming gap functions $d_{xy}$ form factor gives a sign change to the outgoing gap function. The triplet component has an additional chiral $p$-wave factor, which also gives a sign change between incoming and outgoing gap functions, which gets canceled by the sign change from the $d_{xy}$ factor, hence there is no overall sign change. This is qualitatively equivalent to an extended $s\mbox{-}$wave state. For triplet dominant cases, a ZBCP with splitting is the outcome of this lack of sign change. In case of a $s$-wave superconductor, two opposite contributions to the conductivity exist: while the loss of density of states reduces the conductivity, Maki-Thompson like processes result in an enhancement. For isotropic $s$-wave systems, these two  effects cancel at  zero energy. However, the additional anisotropy from the orbital form factor may not give an exact cancellation, this could lead to an increased conductivity at the zero energy in comparison with a pure isotropic $s\mbox{-}$wave superconductor. In the limit of strong singlet component, we find a featureless conductivity. This is expected because the nodal line in parallel to the interface normal, and in such orientation no proximity effect occur\cite{Asano2001}.

\textit{Concluding Remarks--}In this paper, we have studied the conductance of  T-shaped junctions  connected to a superconductor under the influence of a strong Rashba SOC generated by the underlying substrate.
The $\mathbf{d}\mbox{-}$vector in the superconducting state is determined by the SOC. The superconducting state is a mixed parity state with both singlet and triplet components. 
We calculated the tunneling conductance for this system within the quasiclassical formalism.  The effect of the superconducting order is included through Nazarov-Tanaka boundary conditions. We looked at the effect of the device size on the zero-bias conductance peak. In agreement with  earlier work, we find  that smaller device dimensions result in larger FWHM  of the ZBCPs.  Moreover, we showed that both triplet and singlet components affect the conductance. 

Specifically, we considered $sp$, $df$ and $dp$ pairing states. The $sp$ and $df$ states produce ZBCPs, whenever the triplet component is stronger than the singlet one. The peak is weaker in the case of $df$ superconductors due to anisotropic $d_{x^2-y^2}$ orbital form factor. In the strong singlet limit, we find a dip structure in the conductance spectrum. For the $sp\mbox{-}$state,  in the regime where a finite but small triplet component co-exists with a large singlet component, we predict a  weak ZBCP on top of the dip structure. This ZBCP disappears quickly with the decreasing triplet strength. For the $df\mbox{-}$state, the weak ZBCP disappears rapidly already when the triplet component becomes smaller than the singlet component. In contrast, we find a ZBCP splitting for the $dp\mbox{-}$state, which happens because the triplet component does not cause a sign change of the incoming and outgoing gaps. Thus, we conclude that
 making interfaces with different crystallographic orientations of the superconductor will be useful for drawing concrete conclusions in systems, where anisotropic orbital form factors are likely to be present.

In the context of the recent experimental results on \ce{CoSi$_2$}/\ce{TiSi$_2$} heterostructures\cite{Chiu2020}, we believe that the $sp$ state is  consistent with the experiments. We have performed our calculations for device sizes that are comparable to the experimental setup. 
We found that in the triplet dominant regime, the opposite chirality superconductivity on the two helical bands gives a ZBCP in the conductance of the T-junction. This peak is quite robust and stronger than the peak expected for a usual chiral $p\mbox{-}$wave superconductor\cite{Asano2007}. Therefore, we think that the \ce{CoSi$_2$}/\ce{TiSi$_2$} heterostructure is a triplet dominant ($\Delta_{singlet} < \Delta_{triplet}$) superconductor. 
The conductance for an SN junction comprised of such a triplet dominant  mixed parity $sp$ superconductor and a normal metal junction has been studied earlier\cite{Iniotakis2007}, and  agrees  with the \ce{CoSi$_2$}/\ce{TiSi$_2$} tunnel junction data barring the sharp feature at the zero energy.

One of the major issue with our description of the  \ce{CoSi$_2$}/\ce{TiSi$_2$} heterostructures is the lack of the time reversal symmetry breaking (TRSB)  that has been observed up to $T_c$. The TRSB in the mixed parity superconductors has been predicted earlier,\cite{Timm2015,Wang2017} however it is expected to happen at a lower temperature  below $T_c$. Twin boundaries can also cause TRSB, if the triplet and the singlet components are comparable in magnitudes\cite{TB_Sigrist1}. Another possible explanation for the hysteresis observed  in the magnetoresistance data is the Zeeman field induced supercurrent. In a superconductor with broken inversion symmetry, an in-plane Zeeman field gives rise to a supercurrent flow along the direction perpendicular to it\cite{Yip2002,Edelstein2003}. We think that  $\mu$SR experiments on \ce{CoSi$_2$}/\ce{TiSi$_2$} heterostructures will provide an ubiquitous evidence for TRSB.

We have considered a simple one band model for the \ce{CoSi$_2$} for qualitatively understanding the \ce{CoSi$_2$}/\ce{TiSi$_2$} heterostructures. However, it is a multiband system, which can be a possible origin of the TRSB. We leave this issue of TRSB for future study.
We conclude that  the \ce{CoSi$_2$}/\ce{TiSi$_2$} heterostructure is a $s+p$ mixed parity superconducting state with a dominant $p\mbox{-}$wave component. Such  mix parity superconductor with a dominant triplet component is a topologically nontrivial system and is similar to a quantum spin hall system\cite{Nagaosa2009,Sato2009}. It hosts topologically protected Andreev bound states, which carry spin currents, therefore constituting an important platform for further research.

\section*{Acknowledgments}
The authors are grateful to Shao-Pin Chiu and Juhn-Jong Lin for helpful discussions. VM, YL and FCZ are partially supported by NSFC grant 11674278 and by the priority program of the Chinese Academy of Sciences grant No. XDB28000000, and  by the China Postdoctoral Science Foundation under grant No. 2020M670422 (YL).   Work at Zhejiang University  was in part supported by the National Key R\&D Program of the MOST of China, grant No. 2016YFA0300202 and the National Science Foundation of China, grant No. 11774307.
\bibliographystyle{apsrev4-1}

\end{document}